\documentclass[%
reprint,
superscriptaddress,
nofootinbib,
nobibnotes,
 amsmath,amssymb,
 aps,
prl,
longbibliography,
]{revtex4-1}

\usepackage[x11names]{xcolor}
\usepackage{graphicx}
\usepackage{dcolumn}
\usepackage{bm}
\usepackage{float}
\usepackage{tikz-cd}
\usepackage{makecell}
\usepackage{hhline}
\usepackage{hyperref}
\hypersetup{colorlinks, 
	linkcolor={blue!75!black!80!yellow},
	citecolor={blue!75!black!80!yellow}, 
	urlcolor={blue!75!black!80!yellow}
	}

\usepackage[normalem]{ulem}
\usepackage{color}

\usepackage{changes}

\makeatletter \renewcommand\@make@capt@title[2]{%
\@ifx@empty\float@link{\@firstofone}{\expandafter\href\expandafter{\float@link}}%
\sffamily{\textbf{#1}}\@caption@fignum@sep#2 }

\begin{document}


\title{Machine learning band gaps from the electron density}

\author{Javier Robledo Moreno}
\email{jrm874@nyu.edu}
\affiliation{
Center for Computational Quantum Physics, Flatiron Institute, New York, NY 10010 USA
}
\affiliation{Center for Quantum Phenomena, Department of Physics, New York University, 726 Broadway, New York, New York 10003, USA
}

\author{Johannes Flick}
\email{jflick@flatironinstitute.org }
\affiliation{
Center for Computational Quantum Physics, Flatiron Institute, New York, NY 10010 USA
}

\author{Antoine Georges}
\email{ageorges@flatironinstitute.org}
\affiliation{
Center for Computational Quantum Physics, Flatiron Institute, New York, NY 10010 USA
}
\affiliation{Coll{\`e}ge de France, 11 place Marcelin Berthelot, 75005 Paris, France}
\affiliation{CPHT, CNRS, {\'E}cole Polytechnique, IP Paris, F-91128 Palaiseau, France}
\affiliation{DQMP, Universit{\'e} de Gen{\`e}ve, 24 quai Ernest Ansermet, CH-1211 Gen{\`e}ve, Suisse}

\date{\today}

\begin{abstract}
A remarkable consequence of the Hohenberg-Kohn theorem of density functional theory is the existence of an injective map between the electronic density and any observable of the many electron problem in an external potential. In this work,  we study the problem of predicting a particular observable, the band gap of semiconductors and band insulators, 
from the knowledge of the local electronic density. Using state-of-the-art machine learning techniques, we predict the experimental band gaps from computationally inexpensive density functional theory  calculations. We propose a modified Behler-Parrinello (BP)  architecture that greatly improves the model capacity while maintaining the symmetry properties of the BP architecture. Using this scheme, we obtain band gaps at a level of accuracy comparable to those obtained with state-of-the-art and computationally intensive hybrid functionals, 
thus significantly reducing the computational cost of the task.
\end{abstract}

\maketitle

\begin{figure*}[t]
    \includegraphics[width=.95\textwidth]{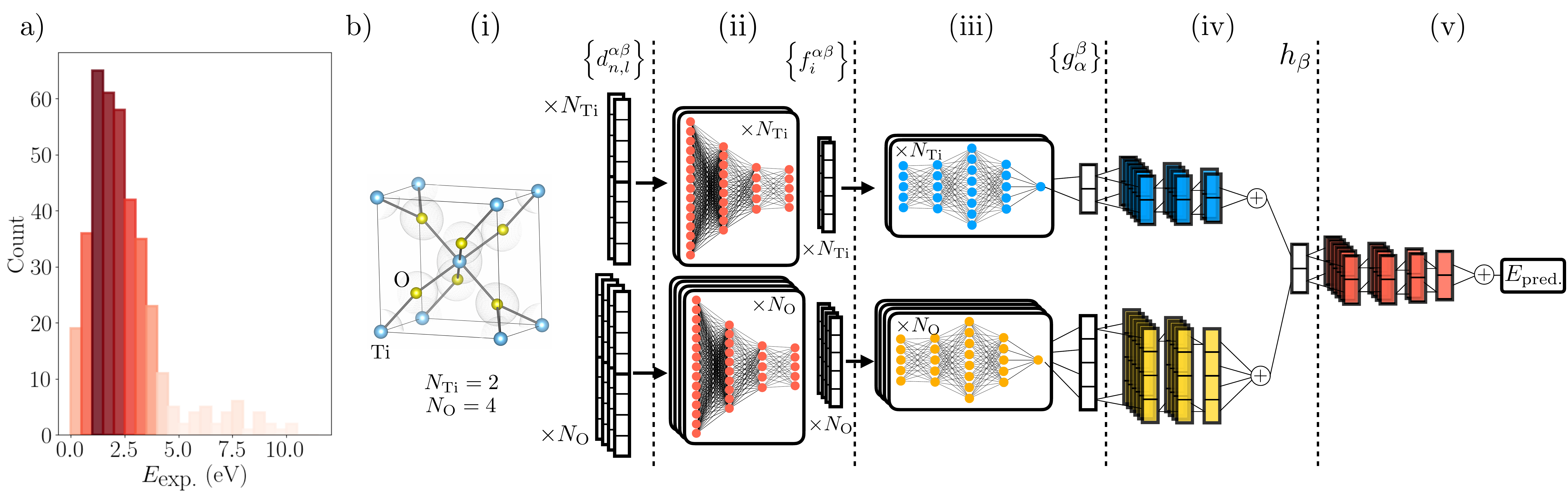}
    \caption{\label{FIG_01: NN-diagram} \textbf{(a)} Histogram of the distribution of experimental band gaps used for training and validation, from Ref.~\cite{Borlido2019}. Histogram box width is $0.5\; \textrm{eV}$. Colors represent the height of each bin, matching the color labels in Fig.~\ref{FIG_03: scatter plots}.
    \textbf{(b)} Diagram of the neural network architecture used to approximate the density to band gap map in the specific example of TiO$_2$. Red modules are shared among all atomic species. Blue and yellow modules are specifically applied to Ti and O densities respectively. From left to right: (i) Rotationally invariant atom-centered density descriptors [see Eq.~\ref{Eq_03: rotationally invariant density descriptors}]; (ii) feature extraction for dimensional reduction; (iii) Behler-Parrinello block and concatenated output $\left\{g^\beta_\alpha \right\}$; (iv) species-specific Deep Sets neural network and concatenation of output of each species $h_\beta$; (v) species-combiner Deep Sets network whose output is the band gap.} 
\end{figure*} 
\section{Introduction}
The Hohenberg-Kohn (HK) theorem of density functional theory (DFT) establishes a one-to-one correspondence between the ground-state local charge density $\rho\left(\vec{r}\right)$ and the external potential of a many-electron system ${v\left(\vec{r}\right)}$~\cite{Hohenberg1964}. Thus, the Hamiltonian of this system is uniquely and unequivocally determined by $\rho\left(\vec{r}\right)$. This remarkable statement implies that any observable of the system must be a functional of the ground-state density. However, the functionals are unknown in most cases~\cite{Cohen2012, KohnNobelLecture}, and in practice physically motivated approximations are used, such as the local density approximation (LDA)~\cite{KohnNobelLecture} or the Perdew–Burke–Ernzerhof (PBE)~\cite{Perdew1985} exchange correlation functionals. Despite the success of these approximations in certain regimes, they can fail to describe relevant observables, such as band gaps of semiconducting materials~\cite{Borlido2020}. 


For modern electronic, optoelectronic, and photovoltaic applications~\cite{Sze2006, Quentin2019}, the reliable estimation of band gaps of solids is of great relevance. The \textit{ab initio} computation of accurate band gaps tends to be a resource-intensive task, as the cheaper exchange correlation functionals like the LDA or generalized gradient approximations (GGAs) underestimate the value of the band gap due to the vanishing of the so called derivative discontinuity~\cite{Borlido2020, Perdew1985, Perdew20017,Dimitrov_2016}. Meta and hybrid GGA functionals, like the Heyd–Scuseria–Ernzerhof (HSE), have a nonzero derivative discontinuity, providing more accurate band gap predictions at a much higher computational cost~\cite{Borlido2019, Borlido2020}. Data-driven approaches have been used to improve the band gap prediction at a lower computational cost. Some works use the PBE band gap predictions as a starting point to compute corrections on the level of HSE or GW accuracy~\cite{Pilania2017, Lee2016}. Another popular approach is to predict the band gap from the positions of the atoms in the unit cell and the crystalline structure~\cite{Na2020, Huang2019, Olsthoorn2019, Dong2019, Pilania2013, Nemnes2019} using synthetic data sets for training. Another common approach is the prediction of band gaps from a set of relevant physical quantities such as boiling and melting points, atomic radii, and bond lengths, among others~\cite{Kauwe2020, Liang2019, Wang2019, Rajan2018, Zhuo2018, Pilania2013}. Previous studies have successfully demonstrated that data-driven approaches can be used to approximate the HK maps between the electron density and observables of interest~~\cite{Brockherde2017, Ryczko2019, Pilati2019, Nelson2019, RobledoMoreno2020}. More recent studies have considered the prediction of accurate band gaps from an approximate description of the electronic density~\cite{Lentz2020, Bruneval2020, Kolb2017}. Despite outstanding progress, most works rely on synthetic training sets containing a relatively small number of different materials in different configurations. 

In this article, we study the problem of inferring experimental band gaps from DFT computed valence electron density distributions in a wide range of bulk semiconductors and band insulators, using deep neural networks (NNs). We propose a Deep Sets~\cite{Zaheer2018} aided modification of the Behler-Parrinello architecture, improving the model capacity while maintaining the permutation invariance in the output. We provide empirical evidence suggesting that learning from the density is advantageous as opposed to learning directly from atomic positions, due to the chemical bonding information explicitly encoded in the electronic density. We find that learning band gaps from PBE-computed densities yields errors comparable to the direct calculation of the band gap using the computationally intensive and accurate HSE06 hybrid functional~\cite{HSE2003}. Finally, we test the neural network performance in a collection of monolayer materials, including examples of the molybdenum family of the transition metal dichalcogenides (TMDCs) and hexagonal boron nitride. 


\begin{table*}[t]\begin{center}
\begin{tabular}{p{4.5cm} | p{12.5cm} } 
   \hline
   
   \hline Name & Description \\ [0.5ex] 
 \hline \hline
 \multicolumn{2}{c}{Operations} \\
 \hline
 $\bm{PP}(x)$  &  Pre-processing fully connected NN for feature extraction.\\ 
 $\bm{BP}^\beta(x)$  &  BP fully connected NN of atomic species $\beta$.\\ 
 $\bm{DS}^\beta(x)$ &  Deep Sets NN of atomic species $\beta$.\\ 
 $\bm{DSsc} (x)$ & Deep Sets NN combining the output of each species block. \\
 $\oplus ^\gamma$ & Concatenation across index $\gamma$. \\
 \hline
 \multicolumn{2}{c}{Data structures} \\
 \hline
 $\left\{d_{n,l}^{\alpha \beta} \right\}$  &  Rotationally invariant density descriptors; see Eq.~\ref{Eq_03: rotationally invariant density descriptors}.\\
 $\left\{f_i^{\alpha \beta} \right\} = \left\{\bm{PP} \left(d_{n,l}^{\alpha \beta}\right) \right\}$ & Pre-processed density features. $i = 1,...,5$.\\
 $\left\{g^\beta_\alpha \right\} = \left\{\oplus ^\alpha \bm{BP}^\beta \left(f_i^{\alpha \beta} \right)\right\}$ & Output of the BP block concatenated for atoms in the same species. \\
 $h_\beta = \oplus ^\beta \bm{DS}^\beta \left(g^\beta_\alpha\right)$ & Concatenation of the output of the species-specific Deep Sets across the different species. \\
 $E_\textrm{pred.} = \bm{DSsc} \left( h_\beta \right) $ & Output of the Deep Sets species combiner. Predicted band gap.\\
 \hline
 
 \hline
\end{tabular}
\end{center}
\caption{\label{Tab0:NN architecture} Detailed description of the modified Behler-Parrinello architecture and the intermediate data structures generated. See Fig.~\ref{FIG_01: NN-diagram}(b) for a schematic representation of the architecture and data structures. Bold symbols label trainable neural networks. In the data structures field, upper indices label the objects on a given set whereas lower indices label elements of an array.}
\end{table*}

\section{Training set and density descriptors}
The data set we use to train and validate our neural network is the set of bulk materials compiled in Ref.~\cite{Borlido2019}. It contains 472 materials for which a reliable and accurate experimental band gap measurement is available in the literature. The reported band gaps are mostly from optical absorption experiments; hence we only consider the determination of direct gaps in this article 
and not of the fundamental (photoemission) or transport gaps. We note furthermore that the reduction of the optical gap by excitonic effects, while relevant in principle when comparing to DFT or HSE determinations of the gaps, is typically only a few tens of millielectronvolts in bulk systems, much smaller than the typical error on the band gaps from these methods. Most materials contain less than 24 atoms per unit cell, while none exceeds 32 atoms. No magnetic materials are included, as the band gap can be highly affected by the magnetic configuration in antiferromagnets. The experimental band gap distribution is shown in Fig.~\ref{FIG_01: NN-diagram}(a). This data set also provides the Materials Project~\cite{MaterialsProject} identification number. This number allows for the extraction of the relevant unit cell parameters to compute the density distribution. 

The density distribution corresponding to the valence electrons for each material is computed from a lattice relaxation \textit{ab initio} DFT calculation using the projector-augmented wave (PAW) method implemented in the Vienna Ab initio Simulation Package (VASP)~\cite{VASP_01, VASP_02, VASP_03}. 
Different levels of approximation are used to compute the density distributions. Specifically, we use LDA and PBE exchange correlation functionals. 
In the latter, we also include van der Waals corrections using the DFT-D2 method ~\cite{Grimme2006}. 

The obtained density values are specified on a uniform grid in the unit cell. Different materials have different unit cell geometries and therefore different grids. Hence, the density value in each grid cell cannot be used as the input of a traditional neural network. Instead, we project the density onto an atomic centered set of orbitals:
\begin{equation}\label{Eq_01: density basis}
    \rho^{\alpha \beta}(\vec{r}-\vec{R}^{\alpha \beta}) \approx \sum_{\substack{0\leq n \leq n_\textrm{max}\\
                  0\leq l \leq l_\textrm{max}\\
                  -l \leq m_l \leq l}}
 C_{n,l,m}^{\alpha \beta} P_n^{\textrm{orth}}(r) Y_{l,m}(\theta, \phi),
\end{equation}
where $Y_{l,m}(\theta, \phi)$ are real spherical harmonics, $P_n^{\textrm{orth}}(r)$ are orthogonal polynomials in the radial coordinate, and $C_{n,l,m}^{\alpha \beta}$ are the expansion coefficients for atom $\alpha$ of atomic species $\beta$ in position $\vec{R}^{\alpha \beta}$, and therefore a set of density descriptors. Recent works have demonstrated this approach to yield accurate density descriptors for machine learning (ML) applications~\cite{Dick2019, Dick2020}. The set $\{ P_n^{\textrm{orth}}(r) \}$ is obtained by the orthogonalization of the polynomial basis:
\begin{equation}\label{Eq_02: radial polynomials}
    P_n(r) = \frac{2n}{r_0^{2n+1}} \cdot \left( r +\frac{r_0}{2n}\right) \cdot \left(r - r_0 \right)^{2n},
\end{equation}
with $r_0$ a cutoff, chosen to be the radius where the value of the density is 5\% of the value of the density at $r = 0$. We follow the standard orthogonalization procedure proposed in Ref.~\cite{Dick2019}. In order to impose rotational symmetry, i.e. ensure that the band gap prediction does not not change if the unit cell is rotated, we choose the set of rotationally invariant features~\cite{Dick2020, Kazhdan2003, Zucchelli2019, Novikov2018}
\begin{equation}\label{Eq_03: rotationally invariant density descriptors}
    d_{n,l}^{\alpha \beta} = \sum_{m = -l}^l \left[ c_{n,l,m}^{\alpha \beta}\right]^2.
\end{equation}
While the above rotationally invariant descriptors are not the only choice to construct invariant features from spherical harmonics~\cite{Zucchelli2019}, they are widely used in ML applications and have been demonstrated to yield accurate results in the context of ML for DFT~\cite{Dick2020}. Even though the directionality of the principal axes of the density distribution is not present in $d_{n,l}^{\alpha \beta}$ as noted in Ref.~\cite{Kazhdan2003}, the density descriptors still preserve the features of the bonding.

\begin{figure*}[t]
    \includegraphics[width=.95\textwidth]{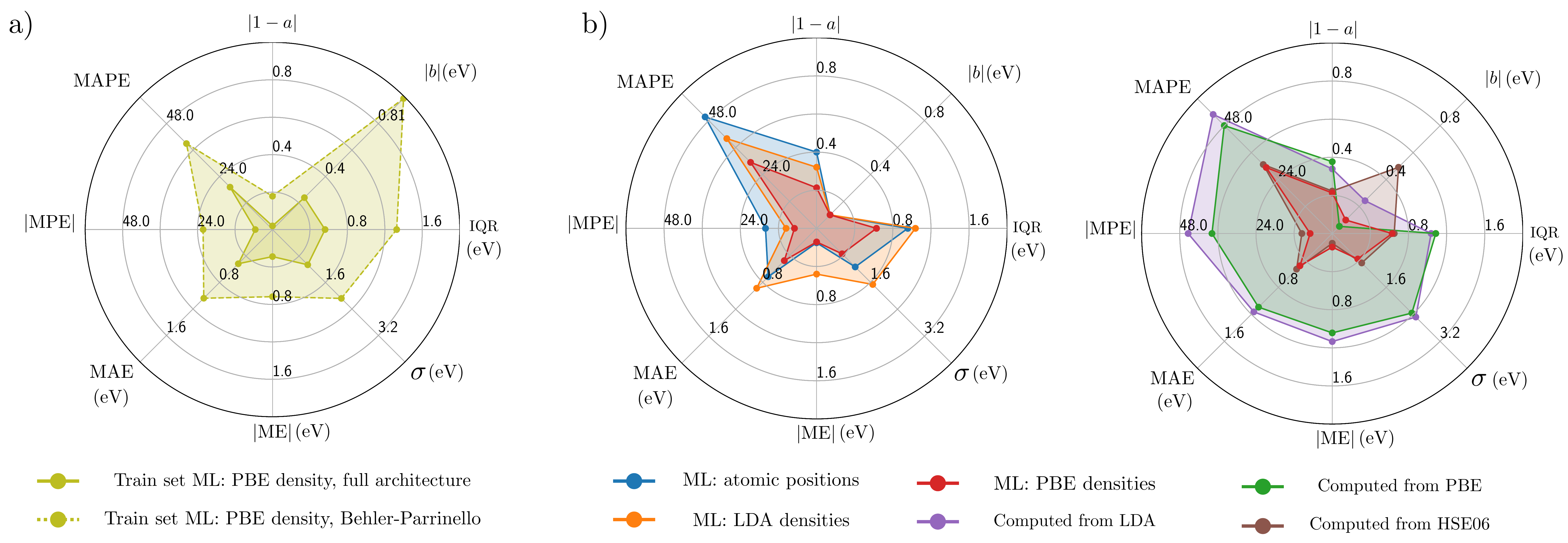}
    \caption{\label{FIG_02: radar plots} Radar plots of the test statistics chosen to measure the band gap prediction accuracy. 
    \textbf{(a)} Train set performance comparison between a Behler-Parrinello architecture--with preprocessing feature extraction--and the full architecture including the Deep Sets modules. Input densities are computed from DFT-PBE.
    \textbf{(b)} Validation set performance in the full architecture including the Deep Sets modules. Left: Band gaps are learned from atomic positions in the unit cell, from densities computed using DFT-LDA and from densities computed using DFT-PBE. Right: Comparison of the best performing ML scheme with DFT-calculated band gaps, using LDA, PBE, and HSE06\cite{Borlido2020} functionals.
        } 
\end{figure*} 

\section{Neural network architecture}
The goal is to infer the value of the experimental band gap given the set of vectors of rotationally invariant density descriptors $\{d_{n,l}^{\alpha \beta} \}$. As a preprocessing step, the set of descriptors $\{d_{n,l}^{\alpha \beta}\}$ are compressed into a set of dimensionally reduced features $\{f_i^{\alpha \beta}: i = 1,...,5 \}$ by a fully connected NN with trainable parameters (the same for all atoms), as depicted in Fig.~\ref{FIG_01: NN-diagram} and Table~\ref{Tab0:NN architecture}.

As noted before, different materials have different numbers of atoms in the unit cell, belonging to different atomic species. In order to have the ability to handle the variable size of input density descriptors, yet respecting the permutation invariance of the input set, we use a modified Behler-Parrinello~\cite{Behler2007} architecture. The NN architecture is depicted in Fig.~\ref{Eq_01: density basis}(b) and detailed in Tab.~\ref{Tab0:NN architecture}. Bold symbols label trainable neural networks. In the data structures field, upper indices label the objects on a given set whereas lower indices label elements of an array. 

The use of Deep Sets~\cite{Zaheer2018} after the BP block ensures the permutation invariance of the predicted gap with respect to permutations of the density descriptors of different atoms, while improving the model capacity by combining in a non-linear manner the outputs of the BP block. Deep Sets layers consist of several channels of the composition of a non-linear activation function and a restricted affine transformation. The restricted affine map imposes permutation equivariance in the output features. It maps an $M$-dimensional input feature $x$ to an $M$-dimensional output feature $y$ in the following form:
\begin{equation}
    y = \left( \lambda \textrm{\textbf{I}} + \gamma \left(\textrm{\textbf{1}} \cdot \textrm{\textbf{1}}^\textrm{T}\right) \right) x + b\textrm{\textbf{1}},
\end{equation}
with $\lambda$, $\gamma$, and $b$ optimizable parameters, $\textrm{\textbf{I}}$ the $M\times M$ identity matrix, and $\textrm{\textbf{1}}$ the $M$-dimensional column vector of ones.

A reduction of over-parametrization effects, owing to the small size of the training set, is achieved via the so called early stopping~\cite{Prechelt2012} and the use of mini batch gradient descent~\cite{Schmidt2019, Lei2018}, with 40 materials per batch.


A collection of statistical measurements of the prediction error is chosen to test the performance of the band gap prediction. Defining the error as the difference between experimental and target band gaps, $E_\textrm{exp.}-E_\textrm{pred.}$, we consider the mean absolute percentage error (MAPE), 
mean percentage error (MPE), mean absolute error (MAE), mean error (ME), the standard deviation of the error distribution ($\sigma$), the interquartile range (IQR), and the parameters ($a$ and $b$) of fitting $E_\textrm{pred.}$ vs $E_\textrm{exp.}$ to $y = ax + b$. We employ stratified 10-fold validation in order to mitigate the possible bias in the error statistics induced by the splitting of the data set in training and validation partitions. Within each partition, we make sure that the distribution of validation band gaps is similar to the band gap distribution of the training set [see Fig.~\ref{FIG_01: NN-diagram}(a)] in order to mitigate error bias. The error statistics correspond to the averages of each statistical measurement over the different partitions.

\begin{figure*}[t]
    \includegraphics[width=.95\textwidth]{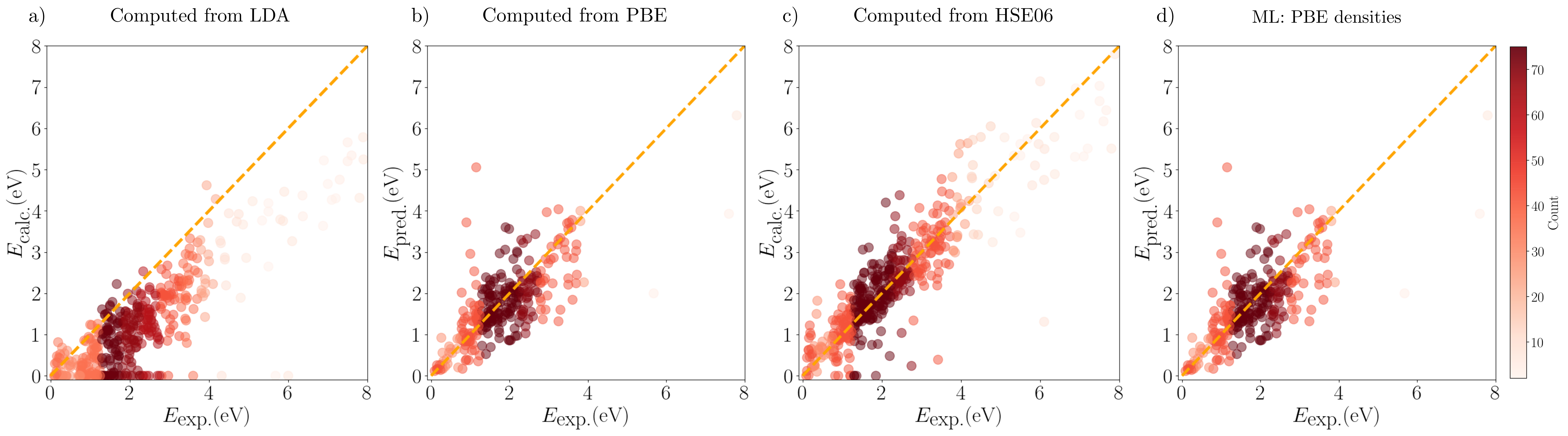}
    \caption{\label{FIG_03: scatter plots} Predicted versus experimental band gaps in the following cases: \textbf{(a)}-\textbf{(c)} Band gaps are computed from DFT calculations using LDA, PBE, and HSE06~\cite{Borlido2020} functionals.
    \textbf{(d)}  ML-predicted experimental band gaps from PBE-computed densities. Only validation set examples are shown. Orange dashed lines are for visual reference, showing what the exact band gap prediction corresponds to.  The color scale for each point corresponds to the number of examples in the training set for that experimental band gap, matching the color scheme of the histogram in Fig.~\ref{FIG_01: NN-diagram}(a). The error statistics of the predicted band gaps shown in this figure are those shown in the right panel of Fig.~\ref{FIG_02: radar plots}(b).} 
\end{figure*} 

\section{Model capacity}
We compare the model capacity of the bare BP architecture  against the Deep Sets improved BP architecture in Fig.~\ref{FIG_02: radar plots}(a). The bare BP architecture does not possess enough expressivity to even interpolate the data. This deficiency in the expressive power stems from lack of  physical justification to take the band gap to be the sum of individual contributions of the different atoms in the unit cell, as done in a bare BP architecture. The improved Deep Sets architecture achieves significantly lower training error by combining in a non-linear and parametrized manner the outputs of the BP architecture. These results show that the addition of the Deep Sets module provides a solution to the model rigidity of the bare BP architecture. This modification, not limited to the problem of band gap prediction, can be implemented in the many applications of the BP architecture.

\section{Learning performance}
The map between the density and physical properties of interest stems from the one-to-one correspondence between the electronic density and the external potential, which fixes the specific form of the Hamiltonian and therefore uniquely determines the whole collection of observables. Obviously, there is also a map from the atomic positions and their nuclear charges to the band gap. 
We compare the prediction of band gaps obtained directly from the atomic positions against the prediction from electronic densities, computed to different levels of approximation (LDA and PBE) in the left panel of Fig.~\ref{FIG_02: radar plots}(b). The test error is larger if atomic positions are used as inputs to the ML model.  This difference is particularly noteworthy in the the MAPE, which is reduced from $\sim 48\%$ in the case of atomic positions to $\sim 28\%$ in the case of PBE densities. The low MPE and ME values indicate that the ML approach does not suffer from a systematic overestimation or underestimation of the band gap. A natural explanation of the prediction improvement arising from the use of the density is that the chemical bonding between the atoms in the unit cell is better taken into account in this approach. The electronic structure of crystals is strongly dependent on the nature of the chemical bonding between their atoms, which is explicitly encoded in the electronic density distribution. Our results thus provide empirical evidence that an explicit encoding of the chemical bonding is beneficial for the task of band gap prediction. This claim is supported by the prediction improvement achieved by using more accurate input densities,  as the neural network trained from PBE densities has better error statistics compared to the model trained from LDA densities. While in some specific cases the LDA provides more accurate ground-state densities, the PBE functional provides more accurate densities on average. 

The right panel of Fig.~\ref{FIG_02: radar plots}(b) shows the error statistics comparison between various DFT approaches to directly compute the band gap and our best performing ML scheme. The DFT-calculated band gaps come from three levels of approximation: LDA, PBE, and HSE06 energy functionals. The LDA and PBE band gaps were obtained from the same DFT calculations used to compute the input densities. The errors are consistent with those in Ref.~\cite{Borlido2020}. The HSE06 band gaps are found in Ref.~\cite{Borlido2020}. The ML-predicted band gaps greatly outperform standard DFT-calculated band gaps in the LDA and PBE approximations. 
 The error in the ML-predicted band gaps is comparable to that of the accurate~\cite{Borlido2020} and computationally intensive HSE06 functional. 
 These results demonstrate that, using the proposed statistical learning setting, it is possible to infer the band gap at HSE06 level of accuracy, while benefiting from the much cheaper PBE estimation of the electronic density.

Panels (a)-(c) of Fig.~\ref{FIG_03: scatter plots} show the values of the DFT-calculated band gaps vs the experimental values for LDA, PBE, and HSE06 functionals. The LDA and PBE functionals systematically underestimate the correct experimental value, providing a biased estimate of the band gap. This observation is also clear in the corresponding values of the MPE and ME shown in Fig.~\ref{FIG_02: radar plots}(b). The HSE06 provides band gap values with significantly smaller error and bias. The ML predicted band gaps in Fig.~\ref{FIG_03: scatter plots}(d) are centered around the correct experimental values, thus providing a low bias estimate. In this case the band gaps show more spread compared to the HSE06 values. The higher spread is related to the small number of available training examples as discussed in the Appendix. The spread is larger for larger band gaps due to the smaller number of training examples with band gaps larger than $4 \textrm{ eV}$ [see Fig~\ref{FIG_01: NN-diagram}(a) and the Appendix].

\section{Extrapolation: monolayer materials}
Lastly, we test the generalization capabilities of the neural network on a collection of monolayer materials, including the molybdenum family of transition metal dichalcogenides and the hexagonal boron nitride. We choose structures with available freestanding experimental band gaps or on substrates with lattice mismatch $< 2\%$, which leads to almost no lattice distortion. Thus, the presence of the substrate is not expected to significantly alter the value of the measured band gap. The training set does not contain monolayer materials. It includes some examples of the family of TMDCs in the bulk: MoSe$_2$, MoS$_2$ and MoTe$_2$.  

\begin{table}\begin{center}
\begin{tabular}{|c | c | c | c |} 
 \cline{2-4}
   \hhline{~|---|} \multicolumn{1}{c |}{} & \makecell{ML from \\ atomic \\ positions}  & \makecell{ML from \\ LDA \\ densities} & \makecell{ML from \\ PBE \\ densities} \\ [0.5ex] 
 \hline
 MoS$_2$~\cite{Klots2014} & $(52\pm17)\%$  & $(15\pm 5)\%$ &  $(7\pm 2)\%$ \\ 
 \hline
 MoSe$_2$~\cite{Choi2017} & $(44\pm15)\%$  & $(40\pm 12)\%$ & $(15\pm 5)\%$ \\
 \hline
 hBN~\cite{Elias2019} & $(65\pm6)\%$ &  $(62\pm 6)\%$ &  $(51\pm 16)\%$ \\
 \hline
\end{tabular}
\end{center}
\caption{\label{Tab1:TMDCs} Generalization error in the monolayer materials. Mean absolute percentage error of the ML band gaps is shown averaged across the 10 training partitions. Error bars correspond to the standard deviation of the MAPE across the 10 training partitions.}
\end{table}

The mean absolute percentage errors of the ML predicted band gaps are shown in Table~\ref{Tab1:TMDCs}. Except for the case of hBN, we obtain accurate predictions of the band gap of these two dimensional structures. Given that $E_\textrm{exp.} = 6.1 \; \textrm{eV} $, the reason behind the lack of accuracy in the hBN is believed to stem from the reduced number of training examples in that range of band gaps (see the Appendix for a detailed discussion). As the ML model is only trained on bulk structures, the accurate predictions in monolayers shows the generalization capabilities of the proposed statistical learning approach. The results in the two-dimensional structures also show the improvement of accuracy as the input density is improved, supporting the claim that better predictions are related to the better explicit description of the chemical bonding in the density distribution.

\section{Conclusion}
In this paper we propose a supervised deep learning approach to predict the experimental band gaps of solids from the knowledge of an estimate of the electron density. This approach is justified by the Hohenberg-Kohn theorem of density functional theory. Owing to the fundamental lack of model capacity of a bare Behler-Parrinello architecture, we introduce a modification to the model architecture based on the utilization of permutation invariant Deep Sets modules. The modified architecture shows a significant improvement in model capacity. This improved architecture can be implemented in other applications where the bare BP architecture is used~\cite{Dick2019, Dick2020}. We also show that the explicit encoding of chemical bonding information in the electron density provides an advantage over learning the experimental band gaps directly from atomic positions. This observation is supported by the prediction improvement when using more accurate estimates of the input densities. The ML-predicted band gaps achieve an accuracy comparable to state-of-the-art DFT-HSE06 estimation of the band gaps, at the much lower computational cost of PBE calculations. Finally, we test the neural network in a collection of monolayer materials, which are not present in the training set, finding good generalization power when the bad gap is in the range $0$-$4$~eV. We believe that the largest source of error in our approach is the small size of experimental data sets available for training (see the Appendix for a detailed discussion).

\section{Acknowledgements}
J.R.M. acknowledges support from the CCQ graduate fellowship in computational quantum physics. The Flatiron Institute is a division of the Simons Foundation. 
We acknowledge useful discussions with Tim Berkelbach and Giuseppe Carleo.


\renewcommand\thefigure{\thesection A\arabic{figure}}    
\setcounter{figure}{0}    
\
\section{Appendix: Effect of the training set size}

In this appendix we explore the effect of the training set size on the performance of the ML approach proposed in the main text. Starting from the full data set, we remove 5\%, 10\%, 15\%, 20\% and 25\% of the examples (randomly selected) and study the validation performance using 10-fold validation as described in the main text. Examples are removed in a nonuniform fashion in order to maintain the same band gap distribution of the original data set [see Fig.~\ref{FIG_01: NN-diagram}(a)].

\begin{figure*}[t]
    \includegraphics[width=.95\textwidth]{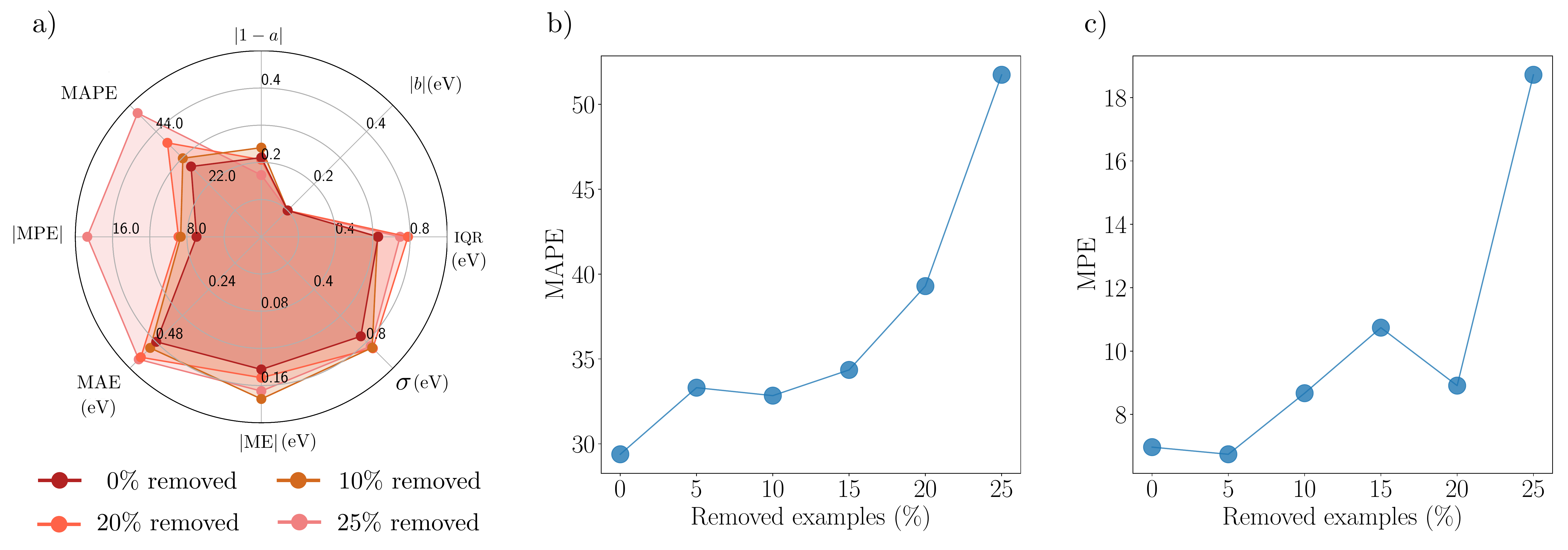}
    \caption{\label{FIG_A_01: train size effect} \textbf{(a)} Radar plots of the test statistics chosen to measure the band gap prediction accuracy. Validation set performance comparison with the data set with removed examples, as indicated by the color labels. \textbf{(b)} Mean absolute percentage error (MAPE) in the validation set as a function of the number of removed examples expressed as a percentage of the total number of examples of the original data set. \textbf{(c)} Mean percentage error (MPE) in the validation set as a function of the number of removed examples expressed as a percentage of the total number of examples of the original data set.} 
\end{figure*}

Fig.~\ref{FIG_A_01: train size effect} shows the data set size effect on the validation error statistics. Panel (a) shows a reduction in the error as the number of materials in the training set is increased. Panels (b) and (c) focus on the MAPE and MPE to quantify the error decrease as the size of the data set is decreased, confirming that the largest source or error of our approach is the small size of experimental data sets available for training.

\begin{figure*}[t]
    \includegraphics[width=.95\textwidth]{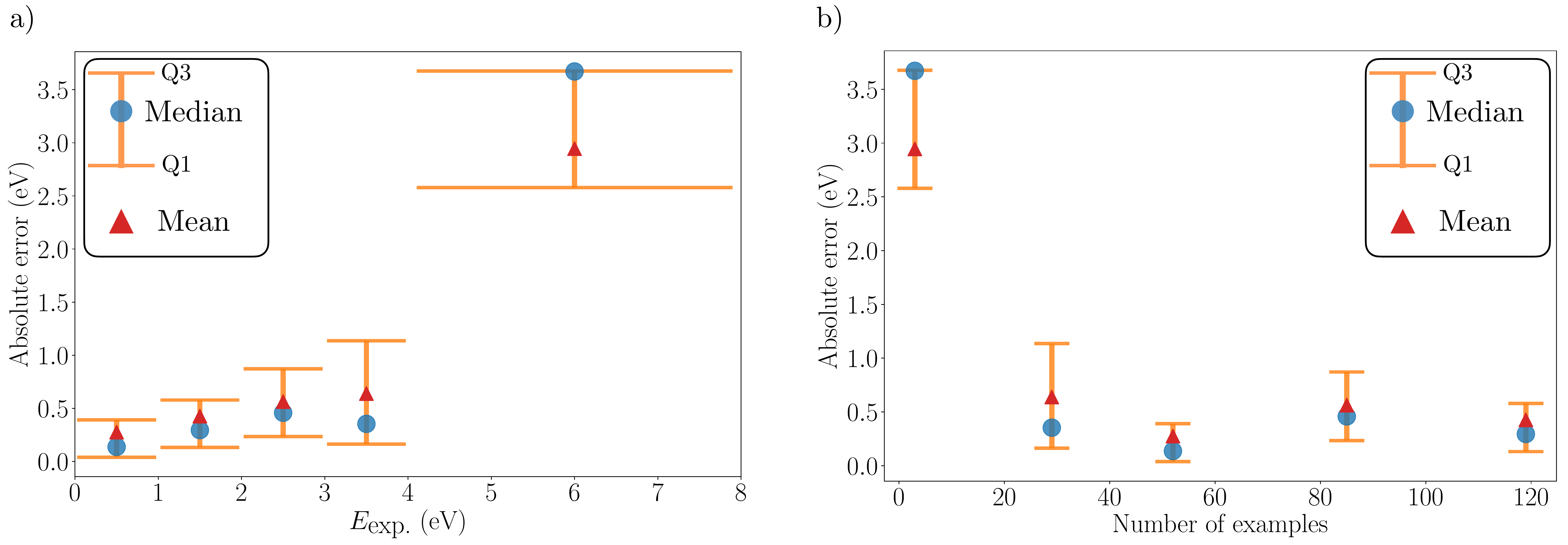}
    \caption{\label{FIG_A_02: effect on number of examples per band gap} \textbf{(a)} Absolute error statistics grouped by band gap value with group separations $[0, 1, 2, 3, 4, 8] \textrm{ (eV)}$ as indicated by the quartile Q1 and Q3 caps. \textbf{(b)} Same as panel (a) with the horizontal axis showing the number of examples in each band gap group. It shows that a small number of training examples in a certain band gap range leads to poor performance.} 
\end{figure*}

Furthermore, we study the effect of the number of training examples in a certain band gap range on the performance of the proposed ML scheme. Fig.~\ref{FIG_A_02: effect on number of examples per band gap}(a) shows the absolute error statistics on different band gap ranges, with separation in $[0, 1, 2, 3, 4, 8] \textrm{ (eV)}$. The error is clearly smaller at smaller band gaps. This error reduction as the band gap is decreased is a direct consequence of having a larger number of training examples with smaller band gap values, as shown in Fig.~\ref{FIG_A_02: effect on number of examples per band gap}(b).

\bibliography{Bib}

\end{document}